\documentclass[12pt,preprint]{aastex}

\font\eightpt=cmr8

\def\Prob   {\ifmmode {{\rm Prob}} \else {Prob} \fi}
\def\arm    {\ifmmode {{\rm arm}} \else {arm} \fi}

\def\twopi  {\ifmmode {2\pi}   \else {$2\pi$}   \fi}

\def\kms    {\ifmmode{{\rm km s}^{-1}}\else{km s$^{-1}$}\fi}

\def\masy {mas~y$^{-1}$}

\def\uas  {\ifmmode {\mu{\rm as}}\else{$\mu$as}\fi}
\def\deg  {\ifmmode {^\circ}\else {$^\circ$}\fi}
\def\porm {\ifmmode {\pm}\else {$\pm$}\fi}

\def\chisqpdf {\ifmmode {\chi^2_{\rm pdf}}\else {$\chi^2_{\rm pdf}$}\fi}
\def\chisq    {\ifmmode {\chi^2}\else {$\chi^2$}\fi}

\def\ZA    {\ifmmode {\theta_{\rm ZA}}\else {$\theta_{\rm ZA}$}\fi}

\def\TECU  {\ifmmode {\rm {\eightpt TECU}} \else {\eightpt TECU} \fi}

\def\etal {et al.~}
\def\eg   {e.g.,~}
\def\ie   {i.e.,~}

\def\d    {\ifmmode {{\rlap{.}}^\circ}\else {${\rlap{.}}^\circ$}\fi}
\def\s    {\ifmmode {{\rlap{.}}^s}\else {${\rlap{.}}^s$}\fi}
\def\as   {\ifmmode {{\rlap{.}}^{''}}\else {${\rlap{.}}^{''}$}\fi}

\newbox\grsign \setbox\grsign=\hbox{$>$} \newdimen\grdimen \grdimen=\ht\grsign
\newbox\laxbox \newbox\gaxbox
\setbox\gaxbox=\hbox{\raise.5ex\hbox{$>$}\llap
     {\lower.5ex\hbox{$\sim$}}}\ht1=\grdimen\dp1=0pt
\setbox\laxbox=\hbox{\raise.5ex\hbox{$<$}\llap
     {\lower.5ex\hbox{$\sim$}}}\ht2=\grdimen\dp2=0pt

\def\lax{\mathrel{\copy\laxbox}}

\def\pa    {\ifmmode {\psi} \else {$\psi$}\fi}
\def\rPpm  {\ifmmode {r_{\Ro,\To}} \else {$r_{Ro,\To}$}\fi}

\def\vlsr  {\ifmmode {v}\else {$v$}\fi}
\def\vhelio{\ifmmode {v_{Helio}}\else {$v_{Helio}$}\fi}
\def\delV  {\ifmmode {\Delta v}\else {$\Delta v$}\fi}
\def\sigV  {\ifmmode {\sigma_v}\else {$\sigma_v$}\fi}

\def\ura   {\ifmmode {\mu_\alpha}\else {$\mu_\alpha$}\fi}
\def\udec  {\ifmmode {\mu_\delta}\else {$\mu_\delta$}\fi}

\def\ul    {\ifmmode {\mu_l}\else {$\mu_l$}\fi}
\def\ub    {\ifmmode {\mu_b}\else {$\mu_b$}\fi}

\def\uml   {\ifmmode {v_{gr}}\else {$v_{gr}$}\fi}
\def\umb   {\ifmmode {v_b}\else {$v_b$}\fi}
\def\vsrad {\ifmmode {v_{rad}}\else {$v_{rad}$}\fi}

\def\upl   {\ifmmode {v^p_{gr}}\else {$v^p_{gr}$}\fi}
\def\upb   {\ifmmode {v^p_b}\else {$v^p_b$}\fi}
\def\vprad {\ifmmode {v^p_{rad}}\else {$v^p_{rad}$}\fi}

\def\Vo    {\ifmmode {V^{Std}_\odot}\else {$V^{Std}_\odot$}\fi}
\def\Uo    {\ifmmode {U^{Std}_\odot}\else {$U^{Std}_\odot$}\fi}
\def\Wo    {\ifmmode {W^{Std}_\odot}\else {$W^{Std}_\odot$}\fi}
\def\VH    {\ifmmode {V^H_\odot}\else {$V^H_\odot$}\fi}
\def\UH    {\ifmmode {U^H_\odot}\else {$U^H_\odot$}\fi}
\def\WH    {\ifmmode {W^H_\odot}\else {$W^H_\odot$}\fi}
\def\V     {\ifmmode {V_\odot}\else {$V_\odot$}\fi}
\def\U     {\ifmmode {U_\odot}\else {$U_\odot$}\fi}
\def\W     {\ifmmode {W_\odot}\else {$W_\odot$}\fi}

\def\Vs    {\ifmmode {V_s}\else {$V_s$}\fi}
\def\Us    {\ifmmode {U_s}\else {$U_s$}\fi}
\def\Ws    {\ifmmode {W_s}\else {$W_s$}\fi}

\def\Vsbar {\ifmmode {\overline{V_s}}\else {$\overline{V_s}$}\fi}
\def\Usbar {\ifmmode {\overline{U_s}}\else {$\overline{U_s}$}\fi}
\def\Wsbar {\ifmmode {\overline{W_s}}\else {$\overline{W_s}$}\fi}

\def\aone  {\ifmmode {a_1}\else {$a_1$}\fi}
\def\atwo  {\ifmmode {a_2}\else {$a_2$}\fi}
\def\athr  {\ifmmode {a_3}\else {$a_3$}\fi}

\def\pars  {\ifmmode{\pi_s}\else{$\pi_s$}\fi}

\def\Ts    {\ifmmode{\Theta_s}\else{$\Theta_s$}\fi}

\def\To    {\ifmmode{\Theta_0}\else{$\Theta_0$}\fi}
\def\T     {\ifmmode{\Theta}\else{$\Theta$}\fi}
\def\Ro    {\ifmmode{R_0}\else{$R_0$}\fi}

\def\Tdot  {\ifmmode{\dot{\Theta}}\else{$\dot{\Theta}$}\fi}
\def\Tddot {\ifmmode{\ddot{\Theta}}\else{$\ddot{\Theta}$}\fi}

\def\lbv     {\ifmmode {(l,b,v)}\else{$(l,b,v)$}\fi}
\def\lv      {\ifmmode {(l,v)}\else{$(l,v)$}\fi}
\def\lvS     {\ifmmode {(l,v)_{\rm src}}\else{$(l,v)_{\rm src}$}\fi}
\def\lbvS    {\ifmmode {(l,b,v)_{\rm src}}\else{$(l,b,v)_{\rm src}$}\fi}
\def\lbvA    {\ifmmode {(l,b,v)_{\rm arm}}\else{$(l,b,v)_{\rm arm}$}\fi}
\def\lbvRBD  {\ifmmode {(l,b,v,R,\beta,d)}\else{$(l,b,v,R,\beta,d)$}\fi}

\def\Nbins   {\ifmmode{N_{\rm bins}}\else{$N_{\rm bins}$}\fi}
\def\DelD    {\ifmmode{\Delta d}\else{$\Delta d$}\fi}

\def\Rtp     {\ifmmode{R_{tp}}\else{$R_{tp}$}\fi}
\def\Vtp     {\ifmmode{V_{tp}}\else{$V_{tp}$}\fi}
\def\Vx      {\ifmmode{V_x}\else{$V_x$}\fi}

\def\Qij     {\ifmmode{(x^j_i,y^j_i)}\else{$(x^j_i,y^j_i)$}\fi}
\def\Qxij    {\ifmmode{x^j_i}\else{$x^j_i$}\fi}
\def\Qyij    {\ifmmode{y^j_i}\else{$y^j_i$}\fi}

\def\Offset  {\ifmmode{(\Theta_x^j,\Theta_y^j)}\else{$(\Theta_x^j,\Theta_y^j)$}\fi}
\def\Offsetx {\ifmmode{\Theta_x^j}\else{$\Theta_x^j$}\fi}
\def\Offsety {\ifmmode{\Theta_y^j}\else{$\Theta_y^j$}\fi}

\def\grad    {\ifmmode{{\partial p}\over{\partial\theta}}\else{${\partial p}\over{\partial\theta}$}\fi}

\slugcomment{2023 January 11}
\shorttitle{Radio Astrometric Techniques} 
\shortauthors{Reid}

\begin{document}

\title{Techniques for Measuring Parallax and Proper Motion with VLBI}   

\author{M. J. Reid\altaffilmark{1}}
\altaffiltext{1}{Center for Astrophysics $\vert$ Harvard \& Smithsonian,
   60 Garden Street, Cambridge, MA 02138, USA}

\begin{abstract}
Astrometry at centimeter wavelengths using Very Long Baseline Interferometry
is approaching accuracies of $\sim$1 \uas\ for the angle 
between a target and a calibrator source separated by $\lax$1$^\circ$ on the sky.
The BeSSeL Survey and the Japanese VERA project are using this to map the
spiral structure of the Milky Way by measuring trigonometric parallaxes
of hundreds of maser sources associated with massive, young stars.
This paper outlines how \uas\ astrometry is done, including details
regarding the scheduling of observations, calibration of data, and measuring
positions.
\end{abstract}


\section{Introduction} \label{sect:introduction}

This paper focuses on the techniques of differential astrometry using
Very Long Baseline Interferometry (VLBI) at centimeter wavelengths.
Here I use the term
differential astrometry to mean measuring the angular difference between a
target source and one or more background calibrator sources nearby in angle.
Target sources are often molecular masers (associated with young stars, red giants,
and extragalactic sources), X-ray binaries, and active galactic nuclei and quasars.
Calibrator sources are usually quasars at sufficient distances that they have
negligible proper motion and, thus, can yield ``absolute'' positions and motions.

Here I will discuss the interferometric phase, $\phi$, as the observable,
in contrast to using interferometric group delay for geodesy and ``whole sky''
astrometry.  Phase delay, $\tau_p$, is
a monochromatic measure of interferometric delay at frequency $\nu$ and is given by
$$\tau_p = {1\over{2\pi}}~{\phi\over\nu}~~,$$ for $\phi$ in radians.
An interferometer phase shift is equivalent to a shift along a sinusoidal interference
pattern, and one cannot discriminate $\phi$ from $\phi\pm n~2\pi$,
where $n$ is an arbitrary integer.  In contrast, group delay is given by
$$\tau_g = {1\over{2\pi}}~{\partial\phi\over\partial\nu}~~,$$
which requires a significant bandwidth to measure.  Thus, group delay 
is the peak of the broadband response, often called a delay or bandwidth pattern
\citep[see, \eg][]{TMS}.  In principle, phase delays can be more
precise than group delays by a factor of $\Delta\nu/\nu$, where $\Delta\nu$ is
the total spanned observing bandwidth (used for group delays) and $\nu$ is the
center observing frequency.  However, since {\it absolute} phase delays are not
usually measurable (owing to $2\pi$ ambiguities), use of phase delay is generally
limited to differential techniques.  Differencing the phase between a target ($T$)
and a calibrator ($C$) nearby on the sky (\ie $\Delta\phi = \phi_T - \phi_C$)
reduces most systematic sources of error by a factor of the angular difference,
$\Delta\theta$ (in radians), between target and calibrator.

For detailed reviews of radio astrometry see \citet{Reid:14} and \citet{Rioja:20}.
Here, instead, I discuss many practical details and lessons learned over the
past several decades.   Section \ref{sect:basics} covers basic requirements
for accurate VLBI astrometry, including some useful rules-of-thumb related to
limiting sources of error.  In Section \ref{sect:observations}, I discuss optimal
strategies for scheduling parallax observations, as well as what calibrations 
are essential for high astrometric accuracy.  Next, Section \ref{sect:calibration}
walks the reader through these calibrations, including examples of problems to watch
for in real data.  Some miscellaneous ``things to consider'' are covered in
Section \ref{sect:consider}, and I conclude with thoughts related to the future
in Section \ref{sect:closing}.

\section{Basics} \label{sect:basics}

Estimating a position (offset) for a point-like source will be limited by random
(thermal) noise and systematics.   In an image, the central position of a Gaussian
brightness distribution can be estimated with a precision no better than about
$0.5~\Theta / {\rm SNR}$, where $\Theta$ is the Gaussian full-width at
half-maximum (FWHM) and
SNR is the peak brightness divided by the image (root-mean-square) noise level
\citep[see Eq. (1) of ][]{Reid:88}.  However, for moderately strong sources,
astrometric accuracy is often not limited by thermal noise.  For example, a 10 mJy
source in a Very Long Baseline Array (VLBA) image with 0.1 mJy noise (i.e., SNR=100) and a
beam $\Theta=1$ mas would be limited to a precision of $5$ \uas.
Even with current calibration techniques and reference sources separated by only
$1\deg$ on the sky, systematic errors from uncompensated interferometric delays,
usually owing to the propagation of the signal through the Earth's atmosphere, can
lead to position errors $\sim10$ \uas.

For signal propagation through the troposphere,
interferometric position error, $\Delta\theta$, scales as follows:
$$\Delta\theta \sim \theta_f \times {c\Delta\tau \over \lambda}~~,\eqno(1)$$
where $\theta_f = \lambda/B$ is the fringe spacing, $\lambda$ is the observing
wavelength, $B$ is baseline length, $c$ is the speed of light, and $\Delta\tau$ is the
propagation delay error.  Essentially, Eq. (1) states that a path delay of one wavelength
shifts a position by one fringe.   Note that for non-dispersive delays
wavelength cancels out, indicating that astrometric accuracy is not dependent
on fringe spacing (angular resolution); instead, accuracy is improved by decreasing
delay errors and/or increasing baseline length.  While this holds for non-dispersive
delays in the neutral atmosphere, it does not hold for dispersive delays from the
ionosphere.

A prerequisite for differential astromety is an accurate model for interferometric
delays.  This involves knowledge of antenna locations, absolute source coordinates,
Earth's orientation parameters, atmospheric propagation delays, clock errors, etc.
Of particular importance is the accuracy of the position of the
source used for phase referencing.  While to first order a position error for a 
source shifts all other sources referenced to it by the
same amount, this is not true to second order in the shift as the interferometer
geometry changes with the position on the sky.   Indeed, a phase 
shift owing to a coordinate error at the position of the reference source
cannot be modeled purely as a position offset at the position of another source,
leading to an undesirable {\it relative} position error as well as degrading
its image \citep{Beasley:95}.  For relative astrometry with VLBI at cm wavelengths with
sources across a couple of degrees on the sky, a reference source error of $\sim100$ mas
can dramatically degrade imaging.  To avoid these problems,
one should ensure that a phase-reference source has a position accurate to
better than a few mas.  If the position is not known to this accuracy prior
to the observations, one can often do a preliminary pass through the correlated
data and measure it against another source with a mas-accurate absolute position.

As the angle between a calibrator and target decreases, astrometric sensitivity to
delay errors also decreases.
Rough rules-of-thumb for achieving $\pm10$ \uas\ relative positional accuracy
for sources separated by $1\deg$ on the sky, include baselines accurate to
$\pm1$ cm, source positions accurate to $\pm1$ mas, and propagation delays
accurate to $\pm0.03$ nsec (corresponding to path delays of $\pm1$ cm).
For many antennas and sources used in VLBI arrays, especially those used
for geodesy and absolute reference-frame astrometry, locations and source
positions of this accuracy are readily available, and the major problem
involves variable atmospheric propagation delays.

Signals delayed when passing through the atmosphere are conveniently described as
occurring in tropospheric and ionospheric layers.  When looking vertically,
tropospheric delays are dominated by dry air, which contributes roughly 200 cm of
path delay (path delay is the signal delay multiplied by the speed of light), and
water vapor, which often leads to between 5 and 20 cm of path delay depending on site,
elevation, and weather \citep[see][]{TMS}.   While the dry air component can be
accurately estimated (from ground-based pressure) and removed in the VLBI
correlator, the variable water-vapor component can only be approximated and
residual vertical path delays of $\pm5-10$ cm are common and can typically
change by $\sim$1 cm hr$^{-1}$.  Astrometric accuracy critically depends
on estimating and removing these residual delays.  Observations using
geodetic blocks (see Section \ref{sect:observations}) can greatly reduce
these path-delay errors.

The ionosphere presents a more difficult calibration problem.  The application
of global models of the total electron content (TEC) above a telescope, while
useful to reduce delay errors, leaves roughly $\pm20$\% of the ionospheric delay
in correlated data \citep{Walker:99}.
Since these delays are dispersive, scaling with observing
frequency as $\nu^{-2}$, they are larger at lower frequencies.   At frequencies
$\lax$10 GHz, they usually become the dominant source of delay error.
For example, at 6.7 GHz (a strong masing transition of methanol), residual
path delays after removing a TEC model are often $\sim$5 cm
(corresponding to 5.6 TECUs, where a $\TECU = 10^{16}~e^-~m^{-2}$).
The most successful method to deal with ionospheric delays is to use the
MultiView technique \citep{Rioja:17}, which uses calibrators surrounding the
target source to solve for and remove the effects of planar ionospheric ``wedges''
(see Section \ref{sect:observations}).

\section{Approaches to Observing} \label{sect:observations}

\begin{figure}[h]
\epsscale{0.50} 
\plotone{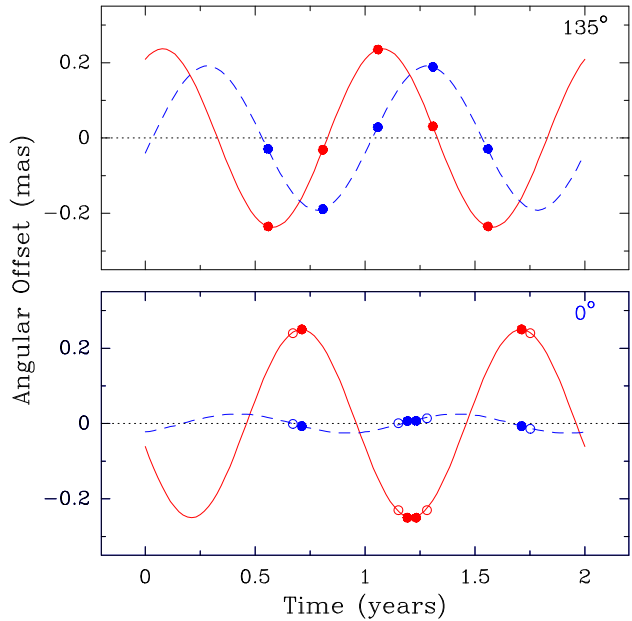}
\caption{\footnotesize Parallax signatures for two sources at 4 kpc distance
  in the Galactic plane:
  (top panel) at $135\deg$ longitude and
  (bottom panel) at $0\deg$ longitude.
  The red solid lines trace the East--West offsets and 
  the blue dashed lines trace the North--South offsets vs.~time of year.
  The plotted points indicate near-optimal sampling.  See the text for
  details.
}
\label{fig:parallax_plot}
\end{figure}

While one would like to avoid phase-unstable weather (e.g., early spring
mornings are generally better than late summer evenings in the southwest U.S.),
one must also take into consideration optimal dates for parallax measurements
when the Earth appears furthest from the Sun as viewed by the source.
(Note that the parallax effect is maximum when the Sun-Earth-Source
alignment traces a right triangle, which implies observations near sunrise and sunset.)
Fig. \ref{fig:parallax_plot} shows the parallax signatures for two locations
in the Galactic plane: at Galactic longitude $135\deg$ in the outer Milky Way
and $0\deg$ toward the Galactic center.  While toward $135\deg$ longitude,
the North--South parallax offsets have nearly the amplitude of the East--West offsets,
this is not the case for the vast majority of sources in the Galactic plane.
Toward the inner Galaxy, the East--West parallax amplitude greatly exceeds that of the
North--South amplitude as exemplified by the $0\deg$ plot.

When the parallax amplitude in the East--West
direction is significantly larger than in the North--South direction, 
one should schedule observations near the extremes of the East--West parallax sinusoid.
An optimal sampling of this sinusoid would be to place one observation
at an extremum, followed by two observations six months later, and then a fourth
one year later than the first.  This scheme decorrelates all parameters in
a parallax and proper motion solution.  However, an undesirable aspect of using
only four observations is that it leaves only one degree-of-freedom
(after solving for offset, motion and parallax) with which to estimate realistic
parameter uncertainties from post-fit residuals \citep[see a detailed discussion in][]{Reid:17}.
Better would be to use eight observations by doubling-up on each of the four described above,
as indicated by the additional open circles in the lower plot of Fig. \ref{fig:parallax_plot}.

The parallax sequences just discussed assume that the sources are detectable
over a full year of observation.  This is usually not an issue for methanol masers,
but can be a problem for water masers.
Assuming water masers come and go on a timescale of seven months,
an optimized  sequence starts at an East--West peak and includes six epochs
at offset days of 0, 91, 155, 189, 241, 364 (or the reverse).
This was determined by evaluating parallax accuracy for thousands of
Monte Carlo simulations, in which the dates of observation were chosen randomly.

A single astrometric observing track should include three types of observations:

\subsection {Fringe finders}
Fringe-finder (a.k.a. manual phase-calibrator) sources are strong,
compact sources which are used to ``find'' interferometric fringes and establish
clock offsets and drifts at each antenna for use in the correlator.
These sources should have mas-accurate positions, so as to avoid residual delays changing
significantly over an observing track, and, ideally, little source structure.
After correlation, one (or more) of these scans are used to estimate and remove
electronic delay and phase differences among separate intermediate frequency (IF)
bands.  Always have at least 2 different fringe-finders observed briefly
(e.g., 1 minute on-source) every couple of hours.

\subsection {Geodetic blocks}
Geodetic blocks typically consist of rapid observations of a dozen
calibrators spread across the sky and observed over a large range of source
zenith angles (\ZA).  The goal is to measure broadband (group) delays and
fit for clock and residual atmospheric path-delays to cm-level precision \citep{Reid:09}.
For a given signal-to-noise ratio, group-delay precision scales inversely
with {\it spanned} bandwidth.
Thus, one should maximize the spanned bandwidth for best precision, even if this
involves gaps in the frequency coverage.  Since fitting delay involves
Fourier transforming frequency to delay, the frequency coverage should
minimize delay sidelobes.  For example, four IF bands,
centered at relative frequencies of 0, 1, 4 and 6 $\times \Delta f$,
yield a ``minimum-redundancy''  sampling of integer multiples of 1 through 6
$\Delta f$.  All calibrators must have positions accurate to $\lax1$ mas,
in order for measured residual delays to be dominated by atmospheric, not
position, errors.  

It is best to observe geodetic blocks at a high frequency
(e.g., 24 GHz), so as to minimize the contribution of the dispersive delay from the
ionosphere.   For dispersive delays, the interferometer phase at frequency $\nu$
is shifted by
$$\phi = -{{c~r_e}\over\nu}~N_e~~,$$ 
where  $c$ is the speed of light, 
$r_e$ is the classical electron radius, and $N_e$ is the electron column density
along the line of sight.  Note that the phase shift is negative.
The phase delay, $\tau_p$, is given by
$$\tau_p = {1\over{2\pi}}~{\phi\over\nu} = -{{c~r_e}\over{2\pi\nu^2}}~N_e~~,$$ 
and the group delay, $\tau_g$, is given by
$$\tau_g = {1\over{2\pi}}{\partial\phi\over\partial\nu} = +{{c~r_e}\over{2\pi\nu^2}}~N_e~~.$$
Note that the group delay has the opposite sign of the phase delay 
(and the radio signal phase shift) for a dispersive delay.
Combining the phase shift and group delay formulae yields the relation
$$\phi = -2\pi~\nu~\tau_g~~.$$  
In other words, when correcting the interferometer phase for a change in
ionospheric group delay, one must flip the sign of the correction compared
to a change in tropospheric delay.
With a mix of dispersive and non-dispersive delays, 
corrections for group delays can be combined, but phase-delays cannot.

If one observes geodetic blocks
at frequencies below about 15 GHz, it is best to measure multi-band delays at
two well-spaced frequencies in order to estimate and remove the dispersive
component of the delay from the total delay, yielding pure non-dispersive delay
for each source.  Note that solving for a zenith delay-error with pure
dispersive delays, and then using an ionospheric mapping function to predict
delays along a ray-path to a source, does not seem to improve astrometric accuracy.
For example, at the high altitude of the ionosphere, rays from a source at, say,
a zenith angle of $45\deg$ will pass through the ionosphere $\approx400$ km
from the antenna site, likely creating a significant azimuthal dependence.
It may be possible to solve for and remove an azimuthal dependence, but this has
not yet been demonstrated.

\subsection{Phase-reference blocks}
Phase-reference blocks involve cycling between sources well within the
interferometer coherence time, which is usually limited by short-term fluctuations
in tropospheric water vapor and produce phase changes proportional to observing
frequency.  For reasonably dry sites, the time spanned
between (the centers of the) phase-reference scans should be less than about
40 s at 43 GHz, 80 s at 22 GHz, and 260 s at 6.7 GHz.  For example,
if a calibrator ($C$) is the phase reference for a target ($T$),
the sequence could be $C, T, C, T,$ ... with scan durations of 40 s at 22 GHz.
Note that scan durations include slew and settle times, so on-source (dwell) times
will be less.  For a strong target and weak calibrator, so-called
{\it inverse} phase referencing can be used: $T, C, T, C,$ ... .  At frequencies
well above 10 GHz, such sequences are adequate, and if more than one
calibrator can be found within about $2\deg$ of the target, they can be used
to give quasi-independent relative positions (provided the calibrators are well
separated on the sky) and to provide a check on potential problems associated
with jet-like structures in some calibrators.

At frequencies below $\sim$10 GHz, MultiView can yield excellent astrometric
accuracy \citep{Rioja:17}.
Ionospheric electron densities follow the Sun, and residual path-delays can be
approximated by ionospheric ``wedges'' and modeled with linear gradients
(a tilted plane) on the sky.  Using calibrators which surround the
target, an observing sequence with four calibrators could be as follows:
$C_1, C_2, T, C_3, C_4, T,$ ... .  This approach involves fitting the phases
from the four calibrators nearest in time to the target for a given interferometer
baseline to a tilted phase-plane:
$$\phi_C = \phi_T + S_x \Delta x + S_y \Delta y~~,\eqno (2)$$
where $\phi_T$ is the desired phase at the target position, $S_x$ and
$S_y$ are phase-gradients in the $x$ and $y$ directions,
and $\Delta x$ and $\Delta y$ are angular offsets from the target position.

In general, three calibrators are the minimum number needed to fit a tilted
phase plane.   In the example above, I mentioned using four calibrators, as 
this provides the opportunity to identify and remove a ``bad'' calibrator,
possibly owing to large phases associated with complex source structures.
Interestingly, however, only two calibrators can be used, provided they nicely
straddle the target source, such that a line between them passes very close
to the target.   

For MultiView to succeed, the entire cycle of calibrators and target must be
completed in well under the interferometer coherence time (e.g. the time it
takes for phases to wander by 1 radian).  For example, at 1.6 GHz, \citet{Rioja:17}
successfully used a cycle time of 300 s.   It is critical that the phases on each
calibrator are dominated by propagation delays and not, for example, by position
errors or source structure.   So, calibrator positions must be known to a small
fraction of the interferometer fringe spacing on the longest baseline.  Note, the
calibrator positions can be updated by measuring relative positions to one of the
calibrators (or possibly the target), which has a well determined absolute position.
Calibrator phases can ``wrap'' past $\pm180\deg$, owing to position or baseline
errors or large residual tropospheric or ionospheric delays in the correlation
process, and these phase wraps must be accounted for when fitting the phase-plane.

If the target source is sufficiently strong to use as the interferometer phase reference,
inverse MultiView has some advantages.   An observing sequence can be
as follows: $T, C_1, T, C_2, T, C_3, T, C_4$, ... , and only the time between
the centers of the target ($T$) scans needs to be less than the interferometer
coherence time.  Also, phase-wrap problems are reduced, compared to standard MultiView,
as atmospheric phase-shifts should largely cancel between the target and each
calibrator.   \citet{Hyland:22} demonstrate that calibrators separated by up
to $7\deg$ from the target can yield excellent results at 8 GHz.  Typically this
allows for many useful calibrators.

\section{Data Calibration} \label{sect:calibration}

\subsection{Geodetic Block Analysis}

The first step in calibration is to estimate clock and residual tropospheric
delays at each antenna.   This can be done with the geodetic-block data.
Start by correcting the data for updated Earth's orientation parameters, antenna
locations, source positions, and feed rotation.   Also, ionospheric delays (usually not
applied during correlation) should be removed, based on total electron content models.
Then, electronic delays and phase offsets among different IFs and feeds should be
``aligned'' using a ``comb'' of frequencies inserted during the observations or
performing a ``manual" phase-calibration using a fringe-finder source.  

Next, multi-band delays for a source on a baseline involving antennas $i$ and $j$,
$\tau_{i,j} = \tau_j - \tau_i~~,$
are fitted with antenna dependent clock parameters and atmospheric zenith delay-errors,
where each antenna's delay is given by
$$\tau = T_0 + {dT\over dt}\Delta T + \tau_0\sec\ZA~~,\eqno(3)$$
assuming other sources of delay error are small.
The first two terms in Eq. (3) correspond to a clock offset, $T_0$,  and a linear clock
drift over time, ${dT\over dt},$ and $\Delta T$ is time relative to a reference
time, best chosen near the center of the observations; the last term is the zenith (vertical)
atmosphreric delay error scaled by the secant of the source zenith angle, $\sec\ZA$, for the
increased (slant) path-length through the troposphere.\footnote{While $\sec\ZA$ is
appropriate for a plane-parallel atmosphere and is
a reasonable approximation for slant path delay, better \ZA\ ``mapping functions''
which account for the Earth's curvature and finite atmospheric thickness
are a significant improvement for observations at large zenith angles;
see \citet{TMS}.}.
Note, with interferometric data the clock parameters for one
(reference) antenna must be held constant and usually are set to zero.
However, since \ZA\ varies differently for each VLBI antenna, one can solve
for the zenith delay at all antennas.
Geodetic blocks with a dozen calibrators can take about 30 minutes to complete,
depending strongly on antenna slew speeds, and should be done every
2--3 hr in order to monitor and correct for changing conditions.

An often overlooked aspect of differential astrometry is that the tropospheric delay
{\it difference} between a target and a calibrator at a given antenna is given by
$$\Delta\tau \approx \tau_0 {\partial \sec\ZA\over\partial\ZA}~~\Delta \ZA 
             = \tau_0~\sec\ZA~\tan\ZA~\Delta\ZA~~.$$
At large zenith angles, both $\sec\ZA$ and $\tan\ZA$ blow up.  While this
can be leveraged to increase geodetic block sensitivity, large \ZA\ observations
should be avoided for phase-referencing astrometry.

\begin{figure}[h]
\epsscale{1.0} 
\plotone{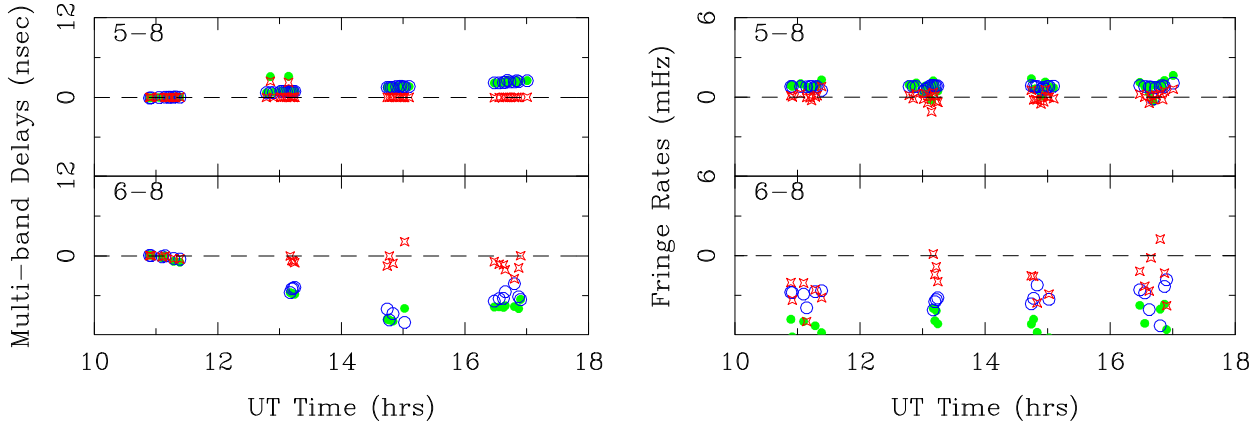}
\caption{\footnotesize Geodetic block examples.  Plotted are multi-band (group)
  delays (left panels) and residual fringe rates (right panels) vs.~time.
  Measurements are green dots, best-fitting model values are blue circles,
  and ``data minus model'' residuals are red stars. 
  The top row is for a well-behaved baseline, whereas the bottom row shows a
  serious problem -- a clock jump between 15:05 and 16:15 UT.}
\label{fig:geodetic}
\end{figure}

Examples of geodetic block data from two baselines of VLBA observations
are shown in Fig. \ref{fig:geodetic}.  Data were taken near 3 and 7 GHz
and the dispersive component of delay (owing to ionospheric electrons)
was calculated and subtracted from each measurement, leaving pure
non-disperive delays.  The left panels plot multi-band (group) delays and the
right panels show fringe rates.  Green dots are the measurements, blue
circles are the fitted model and red symbols are residuals.
The upper plots (for the baseline with antennas 5 and 8) show good data;
one can see a uniform clock drift of about +0.3 nsec hr$^{-1}$ and most residuals
are $\lax0.1$ nsec (or $\lax3$ cm of path delay).   The lower plots
(for the baseline with antennas 6 and 8) show a serious problem.  There
is a large clock-drift rate of $-2.5$ nsec hr$^{-1}$ evident
between 11 and 15 UT (corroborated by an average fringe rate of about $-3$ mHz,
corresponding to $\nu {dT\over dt}$ for observations at $\nu=5$ GHz).
While this is not necessarily a problem, there is a large clock jump between
the third and fourth geodetic blocks (i.e., somewhere between 15:05 and 16:15 UT).
Note that the residuals, while not as small as for baseline 5--8, are reasonably
close to zero.  This can happen as spurious large zenith-delay parameters can
partially compensate for mismodeling the clock as a linear drift.
So, simply looking for small residuals is not sufficient to spot problems.
Clearly the clock jump occurred at antenna 6 (since antenna 8 is common
to both baselines and the jump does not show up in baseline 5--8).  Therefore,
data involving antenna 6 from 15:05 to the end should be flagged,
both in the geodetic block and phase-referencing files, and the geodetic block data
should be re-fitted.

Finally, it is always a good idea to check that the clock and tropospheric delay
corrections work well by applying them to the geodetic block data, re-fitting for
multi-band delays, and verifying that the residual delays are $\lax0.1$ nsec.
Alternatively, since many geodetic block sources are strong, after applying
the corrections, one can simply plot phase versus frequency across all IFs and feeds
and visually verify that the response is nearly flat.

\subsection{Phase-referencing Block Analysis}

The phase-referencing data require standard calibration steps\footnote{see
http://www.aips.nrao.edu/vlbarun.shtml for examples},
with the addition of tropospheric delay calibration using the results
of the geodetic-block fits, and for low-frequency observations
substitution of MultiView for standard phase
referencing.  For example, calibration steps could be as follows:

\begin{enumerate}

\item Flag any bad data.

\item Convert interferometer visibilities (correlation coefficients)
  to Jansky units, using measured system temperatures and antenna gain
  curves, as well as any other known amplitude corrections (e.g.,
  digitization loses).
  
\item Adjust delays and phases as done in the preliminary calibration of
  the geodetic block data (using updated Earth's orientation parameters, 
  antenna locations and source positions, and applying ionospheric TEC models.).

\item Correct phases for antenna feed rotations; for example, for circularly
  polarized feeds shift phases by the feed parallactic angles with opposite
  signs for right and left circularly polarized feeds
  (and be sure to verify the proper sign convention).  Note,
  this holds for linearly polarized feeds correlated against circularly polarized
  feeds, but for linear against linear feeds there is no phase shift induced
  by feed rotation.  The conversion of VLBI data with some or all antennas
  having linearly polarized feeds to a common circularly polarized basis
  is complicated and deserves its own paper
  \citep[see e.g.,][]{Marti-Vidal:16}.

\item Apply clock and tropospheric delay corrections based on fitting
  geodetic block data.  Then remove electronic delays and phases among IFs and
  feeds using either a calibration frequency comb (inserted at the receiver)
  or using a manual phase-calibration on a strong source.  For spectral-line
  sources (e.g., masers) a subtle but serious problem can occur in this step.
  If a second-pass correlation is done which retains only a portion of the 
  total IF bandwidth (so-called ``zoom mode'' used to limit the number of spectral
  channels when high spectral-resolution is desired), then one should also
  correlate the manual phase-calibration source(s) in the same manner and
  apply them to the spectral-line data.  One cannot directly transfer the
  manual phase-calibration from broadband to narrowband data without considering
  the frequency location of the narrowband data relative to the edge frequency of
  its associated broadband, and some software analysis packages (i.e., AIPS and CASA) do
  not have this capability.

\begin{figure}[h]
\epsscale{0.65} 
\plotone{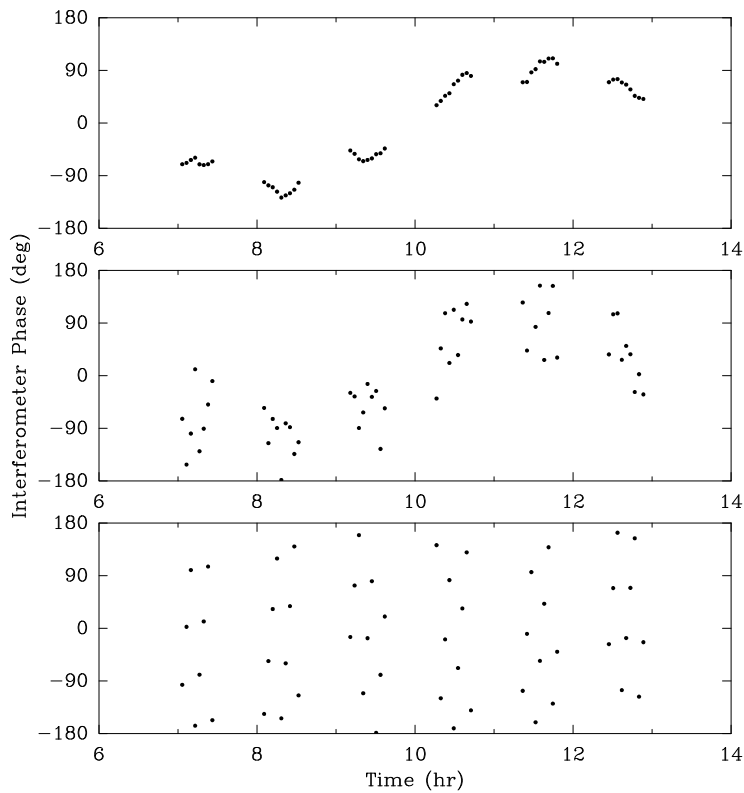}
\caption{\footnotesize Interferometer reference phase simulations.
  Each dot represents a phase measured on a calibrator, to
  be interpolated between two adjacent (closely spaced) measurements
  and removed from the target source.  A strong target can be used as the
  phase reference for a weak calibrator, so-called
  ``inverse phase referencing,'' if desired.  Top panel: high signal-to-noise
  phases showing small variations over $\sim5$ minutes and larger
  variations over hours.  For high-quality differential astrometry,
  reference phases should resemble these.
  Middle panel: the same phases but generated with low signal-to-noise and
  these are far from optimal for phase referencing.  Bottom panel:
  high signal-to-noise phases, but with a high residual fringe-rate as evidenced by
  rapid ``phase wrapping.''  These indicate a large position error for the
  phase-reference source, or other geometric or clock errors, and are not
  suitable for accurate differential astrometry.
}
\label{fig:phases}
\end{figure}

\item If the local oscillator was held constant during the observing track,
  then spectral-line sources will shift in frequency owing to the Earth's
  rotation and orbit about the Sun.   Then interferometer visibilities should be
  corrected to hold a spectral line steady in frequency.  If the visibilities
  are in the delay-lag domain, $V(\tau)$, this can be accomplished by an appropriate
  linear shift of visibility phase as a function of delay-lag, which corresponds
  to a shift in spectral frequency (by the Fourier transform shift theorem).
  If, instead, the visibilities are in the frequency domain, $V(\nu)$,
  one can Fourier transform to delay-lag, perform the linear phase
  shift, and then inverse Fourier transform back to frequency.

\item Phase reference the data.  This can use a single source, either a calibrator
  (standard phase-referencing) or the target (inverse phase-referencing),
  or it can involve multiple calibrators (MultiView).
  Fig. \ref{fig:phases} shows three simulated examples of reference phases
  calculated from a single source.  The top panel shows high SNR phases which
  might be obtained at $\sim$1 cm wavelength.  The slow phase wander of hundreds 
  of degrees of phase over hours is typical of path delays of $\sim$1 wavelength 
  owing to variable water vapor.  Within groups spanning about 15 minutes, the phases
  look ``worm-like,'' which comes from small-scale fluctuations in water vapor,
  and one can reliably interpolate between individual points where the source to
  be referenced was observed.

  The middle panel of Fig. \ref{fig:phases} shows the same simulated data,
  except a low SNR of unity was used for each point.  Interpolation between adjacent
  points would be very inaccurate and, when applied to another source, 
  imaging would be considerably degraded.  The bottom panel shows the effects of
  a large error in the position of the reference source, as evidenced by the
  ``patterned'' appearance made by rapid phase wrapping.  Other errors, such
  as incorrect baselines or uncompensated clock drifts, can also do this.
  In this case, differential astrometric accuracy would be degraded by second-order
  effects of phase referencing as discussed earlier.
  
\begin{figure}[h]
\epsscale{0.71} 
\plotone{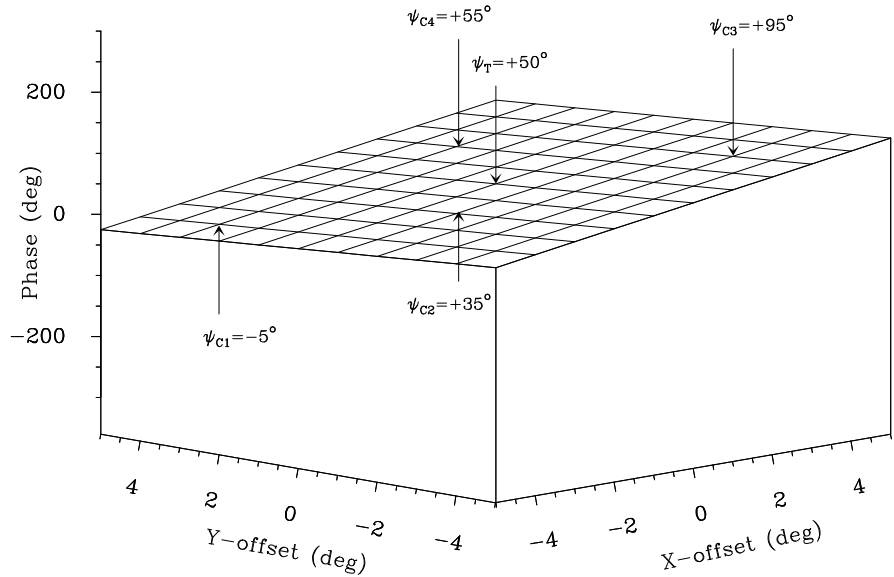}
\caption{\footnotesize MultiView phase-calibration simulation of an
  ionospheric wedge with gradients of $10\deg$ and $-5\deg$ of phase
  per degree of angular offset in the $X-$ and $Y-$directions.  Calibration
  sources are offset from the target at $(X,Y) = (0\deg,0\deg)$ by   
  ($-4\deg,+3\deg$) for C1, ($-2\deg,-1\deg$) for C2, ($+3\deg,-3\deg$) for C3,
  and ($+2\deg,+3\deg$) for C4.
  Fitting a tilted plane to the phases of the four calibrators
  ($\psi_{C1}, \psi_{C2}, \psi_{C3},~{\rm and}~\psi_{C4}$) returns
  the $50\deg$ phase at the origin to be removed from the target source.
  In general, the minimum number of calibrators needed to define the phase 
  plane is three, but it can be useful to use four to test for a ``bad''
  calibrator (e.g., with source-structure phase).  In cases where
  the line between two calibrators comes very close to the target source,
  one can use only two calibrators.
  }
\label{fig:multiview}
\end{figure}

As outlined in Section \ref{sect:observations}, at frequencies below $\sim10$ GHz,
path-delays from ionospheric ``wedges'' can be removed from data with the
MultiView calibration technique.   Fig. \ref{fig:multiview} shows simulated
phases on one baseline at an instant in time for four calibrators surrounding a target.
A tilted plane fitted
to the calibrator phases as a function of $(X,Y)$-offset yields a phase of
$50\deg$ at the position of the target source, which could then be removed 
from the target source data prior to astrometric measurements.   

As discussed in Section \ref{sect:observations}, phase wraps can be a serious problem.
Fig. \ref{fig:phase_wrap} gives a simple example of a phase-wrap issue. The phase for
the C3 calibrator, coming from the $\arctan(V_{imag},V_{real})$ of an interferometer
visibility, is generally defined between $-180\deg$ and $+180\deg$ and, in this example,
is $-175\deg$.   However, this phase should be $+185\deg$ (i.e., $-175\deg+360\deg$)
and must be set to that value when fitting a MultiView tilted-plane to all calibrator
phases.  Otherwise, MultiView would return a spurious phase to the target.

\begin{figure}[ht]
\epsscale{0.71} 
\plotone{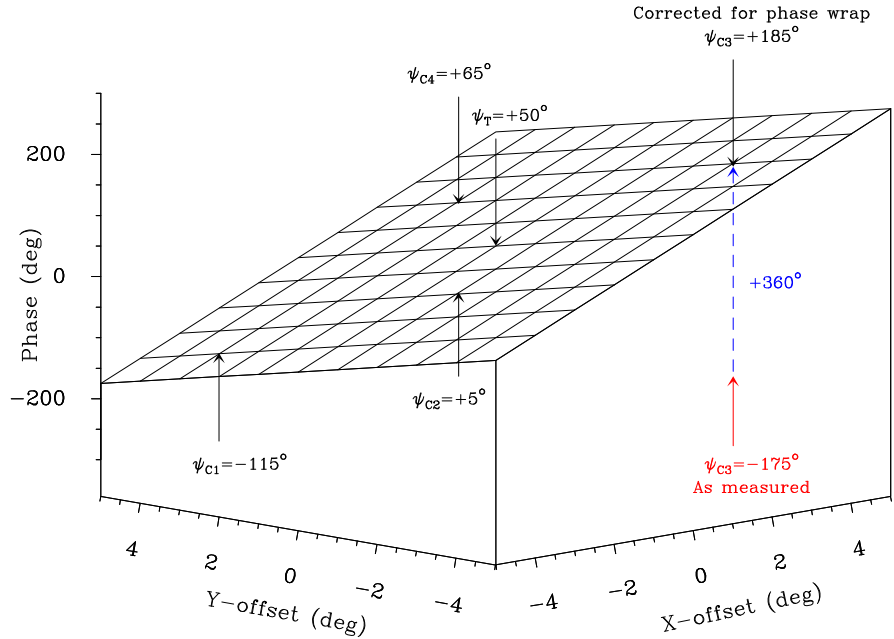}
\caption{\footnotesize MultiView phase-calibration simulation, similar to that in
  Fig. \ref{fig:multiview}, but with larger gradients of $30\deg$ and $-15\deg$ of phase
  per degree of angular offset in the $X-$ and $Y-$directions.   
  Fitting a tilted plane to the phases of the four calibrators
  ($\psi_{C1}, \psi_{C2}, \psi_{C3},~{\rm and}~\psi_{C4}$) would also yield
  the $50\deg$ phase at the origin to be removed from the target source.
  However, with the larger phase slope, the phase for C3 is now $185\deg$,
  which would generally be reported as $-175\deg$.  Fitting a tilted plane using
  that phase would result in a very large error in the slope and in the fitted phase
  at the target position.  Such phase-wraps must be corrected for MultiView to work.
  }
\label{fig:phase_wrap}
\end{figure}

If the target source is sufficiently strong and compact so as to yield a high SNR 
in a single scan, it can be used as a ``preliminary'' phase
reference for the surrounding calibrators (so-called inverse MultiView).
This has the important advantage of
pre-calibrating the data on a shorter time-scale than standard MultiView,
which requires cycling around all calibrators and the target well within the
interferometer coherence-time limitation.  This pre-calibration also results in
``differenced phases,'' which can diminish phase-wrap problems.   The next
step in inverse MultiView is the same as standard MultiView, i.e. fitting a
tilted plane to the differenced phases of the calibrators and subtracting the
phase offset at the target position from the target phases (which were zero after
the pre-calibration step).  This returns ``clean'' phases, which give the desired
position information for the target.  \citet{Hyland:22} have demonstrated that
inverse MultiView can work well for 8 GHz observations and achieve
$\approx20$ \uas\ (single-epoch) astrometry.
  
It is essential that the calibrators used for MultiView have position
errors small enough so as not to contribute significantly to the measured
phases.  If {\it a priori} positions are only known to, say, $\pm0.5\theta_f$, then
measured phases could have $180\deg$ unwanted contribution from their position errors.
For a fringe spacing $\theta_f \sim 1$ mas, this corresponds to a position
error of only 0.5 mas, and not all calibrators have this positional accuracy.
However, what is more important is {\it relative} position accuracy among
the MultiView calibrators.  Relative position accuracy better than
$\pm0.1\theta_f$ can usually be achieved through a first-pass of phase
referencing and imaging of all sources relative to a single source with
a very accurate position.  After correcting the interferometer visibilities
for the measured position offsets, one can then repeat the calibration
sequence, now with greatly reduced contributions to phases
from poor positions, and then do the MultiView step.

\end{enumerate}

\section{Things to Consider} \label{sect:consider}

A cardinal rule for astrometry used to estimate parallax and/or proper motion
is to avoid changing any observational parameters over the entire program.  This 
offers the best chance for canceling systematic errors.  For example, if a source
has structure within the resolution of the array and either the source or
the interferometer $(u,v)$-coverage changes between observations, then one can get
centroid shifts that are unrelated to parallax or proper motion and
can degrade their measurement.

For sources with modest structure, such as a double with separation comparable
to the interferometer resolution, accurate astrometry can usually benefit from
some degree of ``super-resolution.''  Often when imaging, one can use a round
CLEAN restoring beam even if the dirty beam FWHM is elliptical.  I have found it
useful to separate compact emission components by setting the CLEAN restoring beam to
$$\theta_{FWHM} = S_r \sqrt{\theta_{maj}\times\theta_{min}}~~,$$
where the $\theta_{maj}$ and $\theta_{min}$ are the ``dirty'' beam's
major and minor FWHM sizes, and using a moderate super-resolution factor, $S_r$,
between 0.5 and 1.0. 

While one can consider directly fitting interferometer phases for position offsets,
it seems advisable to always make images in order to assess whether or not a source
has a complicated structure.  If a source (either the target, which might be an
X-ray binary, or a quasar calibrator) has a jet-like structure, which changes among
observing epochs, one should consider rotating the measured positions and the
parallax/proper motion model to be along and perpendicular to the jet direction
\citep[\eg see][]{Miller-Jones:21}.
This allows one to down-weight the ``corrupted'' data in the jet direction and rely
more on the ``clean'' data in the perpendicular direction.
  
When measuring group delays, e.g. for geodetic block observations, one should
spread the spanned bandwidth in order to get more accurate delays.  On the
flip side, for phase-reference blocks, it is better to pack the bands together
in order to minimize the frequency spanned and have less sensitivity to delay errors.
Note that modern digital base-band converters (DBBCs) may introduce delay jumps
when changing between different setups, and this can significantly degrade
(or destroy) astrometry.  If one's DBBCs have such problems, it might be
best to observe geodetic and phase-reference blocks with the same electronic setup.

Calibrators offset from a target in decl.~are often not as good for
astrometry as those offset in R.A., since phases associated with atmospheric 
(zenith-angle dependent) delay errors can more closely mimic decl.~than
R.A. offsets.

\section{Closing Thoughts} \label{sect:closing}

The most accurate parallax measurements to date have uncertainties near $\pm5$ \uas\ 
\citep[see Table 1 of][and references therein]{Reid:19}.
In many cases, these come from measurements at 22 GHz and are limited by uncompensated
tropospheric delays.  Since these are likely to be quasi-random, parallax measurement
can be improved with a greater number of observations ($N$), with precision scaling
as $1/N^{1/2}$ (if optimally sampled).
Currently, it is not clear if the non-dispersive phase-delays
associated with fluctuations in water vapor are uncorrelated over several degrees
on the sky or, instead, have a ``planar'' structure which could be addressed
by MultiView calibration.

Three-dimensional (3D) kinematic distances offer a promising method of estimating distances
to sources well past the Galactic center, where observed motions can approach 470 \kms,
\ie twice the rotation speed of the Galaxy \citep{Reid:22}.
Thus, even at a distance of 20 kpc, this leads to a proper motion of about 5 \masy,
or over one year about 100 times the magnitude of the parallax amplitude.
For young stars hosting masers, non-circular motions are $\sim$10 \kms,
and hence contribute less than a couple of percent to the distance error budget.
Usually, proper motion can be measured much more easily than parallax, in part because
motion precision scales as $1/N^{3/2}$ (for uniform sampling in time).  Note, that if
one observes a source at integer year intervals, the parallax effect vanishes,
simplifying the motion measurement.  Measurements with future VLBI arrays, including
the SKA and the ngVLA, may be able to obtain hundreds-to-thousands of 3D kinematic
distances for very distant Galactic sources.

\end{document}